\documentclass{iaus}
\usepackage{psfig}

\title[ Mass distribution] %% give here short title %%
{ The mass distribution in Spiral galaxies}

\author[Salucci]   %% give here short author list %%
{Paolo Salucci  }

\affiliation {SISSA 
  \break Via Beirut, 2-4, 34100 Trieste, Italy  \break email:  salucci@sissa.it}
  
  \pubyear{2007}
\volume{244}  %% insert here IAU Symposium No.
\pagerange{63-72}
\date{?? and in revised form ??}
\jname{Dark Galaxies and Lost Baryons}
   \editors{J. I. Davies
\& M. J. Disney, eds.}

\begin{document}

\maketitle

\begin{abstract}

 In the past years a wealth of observations has allowed us to unravel the structural properties of the Dark and Luminous 
 mass distribution in spirals. As result,  it has been found that  their rotation curves follow, out their virial radius,  an Universal function  (URC)  made by two terms: one due to the gravitational potential of    a Freeman stellar disk and the other due to that of  a dark halo.   The   importance of the latter is found to   decrease  with  galaxy mass.  Individual objects reveal in detail  that dark halos have a density core, whose size correlates with its central  value.
These properties   will guide   $\Lambda$CDM Cosmology to evolve to match the  challenge that observations presently pose.
\end{abstract}

\firstsection % if your document starts with a section,
\section{Introduction}

Rotation curves (hereafter RC) of disk galaxies do not show any
Keplerian fall-off and do not match the distribution of the stellar
(plus gaseous) matter. As a most natural explanation, this implies
an additional invisible mass component (Rubin et al. 1980; Bosma
1981, Persic \& Salucci, 1988) that becomes progressively more
conspicuous  at outer radii and for the less luminous galaxies (e.g.: Persic \& Salucci, 
1988; Broeils,  1992). The  distribution of matter  in disk systems has become a crucial benchmark   for the present 
understanding of   the process of   galaxy formation. Time is ripe that we  can  address,  with the help of  the available observational 
scenario, crucial questions:   i) has the DM    an  universal distribution  reflecting  its very  Nature?  
ii) how and why  the dark-to-luminous mass ratio and  other dark and luminous physical  quantities vary   in objects of different mass?          

 \begin{figure}
\centerline{\psfig{figure=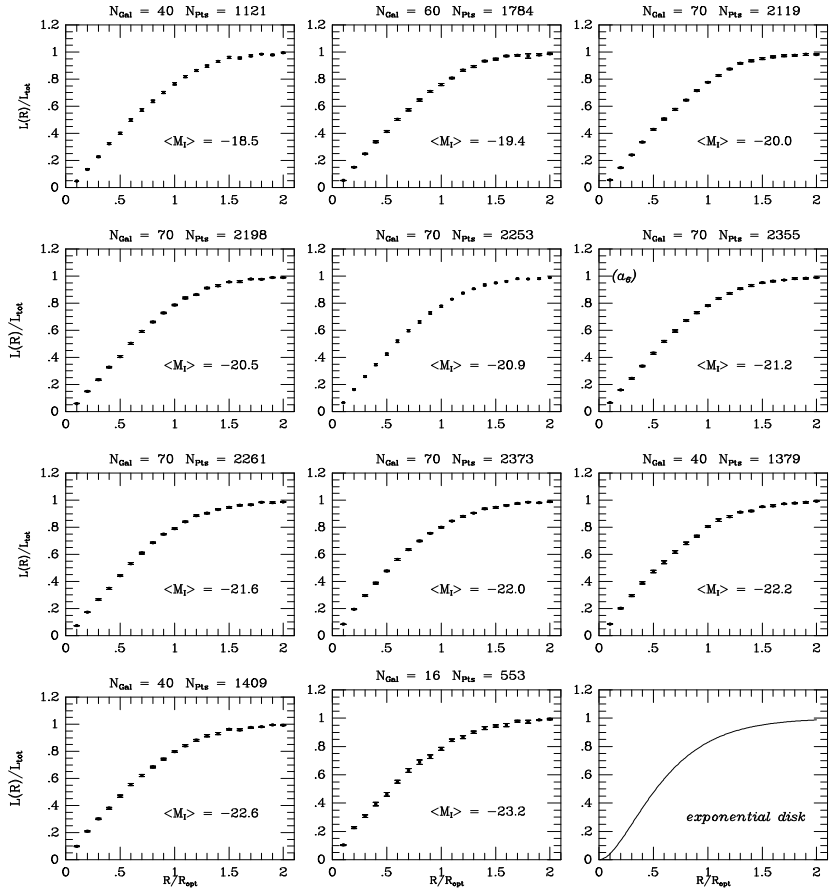,height=11.9cm,width=11.9cm}}
\vskip 0.2cm
\caption{  The  coadded luminosity profiles   $L(R/R_{opt})/L_{tot}  = \int_0^{R/R_{opt}} I(r^\prime)\,r^\prime \,dr^ \prime/L_{tot}$   } 
\vskip -0.2cm
 
\end{figure}

It is well known that numerical simulations performed in the framework of the currently favored theory    
of structure formation,    $\Lambda$ Cold Dark Matter,    
predict a well-defined density profile for the     virialized halos (Navarro, Frenk and White, 1996).
 Furthermore, the mechanism of galaxy formation, as currently understood, involves
the cooling and  the condensation of  pristine  HI gas    inside the gravitational
potential well of  DM halos,    the transformation of part of this gas  into stars  and finally  the feedback of the latter on the former, through   SN explosions.   It is widely accepted that the mass distribution of spiral galaxies, we can  derive from   observations,  bears the imprint of the Nature and  the cosmological history of  the DM, its interaction with the LM 
and it highlights the astrophysical processes that have turned (some of) the pristine infalling  gas into a stellar disk.

It is worth  to define the Dark Matter in the following way: let us set  $M(R)$  the mass  distribution of the  
gravitating matter and $M_L(R)$ that of the baryonic "luminous"  matter,   both obtained from observations.
Realized that  these  distributions do not match, i.e. $ dlog\  M(r)/dlog\  R> dlog\  M_L/dlog \ R$,  we introduce a dark mass  
component with mass profile $M_h(R)$ such that:

 \begin{equation}
{dlog M(R)\over {dlog\  R }} =   {M_L(R)\over {M(R)}} { dlog M_L\over {dlog \ R}} + {M_h(R)\over{M(R)}} { dlog \   M_h\over {dlog \ R}}
  \end{equation} 
 
With this  definition it is immediate that the study of the  DM phenomenon   involves the {\it slope } of the RC rather than its amplitude and must start from a proper knowledge  of the luminous matter distribution. In spirals, this has three components: a stellar bulge,  a thin stellar  disk and an extended thin HI disk.   Within   the aim of this paper the latter component  plays little role and  it will be neglected  unless we consider individual HI-rich objects.   The first component, instead,  it is important in early type  spirals, not considered here  and then we redirect the reader to  Noordermeer et al. (2007). 
 
The  luminous  matter in spirals is distributed in   thin disk with surface luminosity   (Freeman, 1970): 
\begin{equation} 
I(R) ~=~ I_0 ~ e^{-R/R_D}
\label{eq: FREEMAN} 
\end{equation} 
with $R_D$  the disk scale-length;  $I_0$ the central value;  it is useful to define
 $R_{opt} \equiv 3.2\,R_D$ as the "size" of the stellar disk, whose  luminosity  is $L_{tot}= 2 \pi I_0 R^2_D$.
 In Fig. 1 we  plot the coadded (I-band) light profiles
  of 616 late types spirals  arranged in 11 luminosity bins as a function of radius, expressed in units of optical radius. We realize
  that  the light profile  does not  depend
  on galaxy  luminosity and  it is  well represented by a Freeman profile.   
  
  Although the  mass modelling of some individual object may require   to consider
 non-exponential stellar disks,  eq (1.2)  well represents  the typical  distribution of stars in late type spirals. Similarly, 
 the moderate radial variations  of the stellar   mass-to-light ratios in spirals  will be neglected  as they are 
 irrelevant for the present issues (Portinari \& Salucci, 2007).

 \begin{figure} 
 \vskip -0.7cm
\centerline{\psfig{figure= 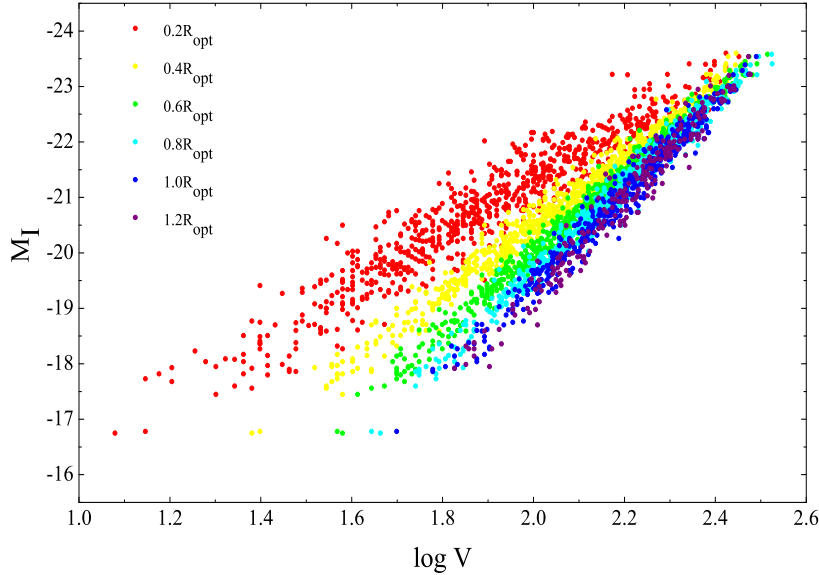,height=9.5cm,width=12.4cm}}
\par
\vskip -0.7cm
\caption{The Radial TF relation} 
\end{figure}

Finally we assume   flat cosmology with matter
density parameter $\Omega_M = 0.27$ and Hubble constant $H_0 = 71~
\mathrm{km~s}^{-1}~\mathrm{Mpc}^{-1}$, at the present time,
   the size $R_{vir}$ and the mass   $M_{vir}$ of a virialized   halo are related by:

\begin{equation}
R_{vir} = 259 \Big( {M_{vir}\over {10^{12} \ \rm M_\odot}}\Big)^{1/3} \ \rm kpc~ 
\end{equation}

\section{The RC's as gravitational field tracer}

In  self-gravitating  systems  in  full  centrifugal equilibrium,  the  circular velocity $V$  at a radius $R$   is linked  to the  gravitational field  $\Phi$ by  $V^2/R=R d\Phi/dR$. In spirals,  we measure the rotation velocity $V_{rot} $,  i.e. the projection on the line of sight of the tangential velocity in cylindric coodinates. Let us notice that:  $V_{rot} =V+V_{na}$, where the latter 
defines non-axysimmetric motions  unrelated to the central potential.    $V_{rot} $ measures the gravitational field only 
when the latter motions are negligible. The     discovery, at radii  $R_n\equiv (n/5)\  R_{opt}$,  of    the    Radial Tully-Fisher relation (Yegorova \& Salucci, 2007), i.e.  of  an  ensemble  of  TF-like, statistically  independent,  low scatter  relationships  between   the  galaxy  absolute  magnitude  and its  {\it rotation}    velocity (see Fig. 2): 
\begin{equation}
M_{band} = a_n \log V_n + b_n\,,
\end{equation}
 where    $V_n \equiv V_{rot}(R_n)$, and   $a_n$,  $b_n$   are the slope and zero-point of the relations,    
indicates  that the RC's are good  gravitational field tracer. More specifically,   the fact that  in any object  and at any radius,   the rotation velocity can be predicted  just by the galaxy luminosity implies that  $V_{na}$ is generally negligible.  
A  similar conclusion is reached  when we  reproduce  the   innermost regions of spirals  RC's with  mass models with   the  luminous matter alone,  that leaves  no space  both for   DM and  for  significant  non circular motions (Ratnam et al., 2000).

\section {The Universal Rotation Curve}       
 
The kinematical properties of Sb-Im spirals led to   the {\it Universal Rotation
Curve (URC) paradigm},   pioneered in Persic and
Salucci (1991) and set in  Persic, Salucci \& Stel (1996, hereafter
PSS). We can state that (see also Salucci et al, 2007) the
 RCs can be generally represented out to their virial radius  by  $V_{URC}(R; P)$, i.e.
by a {\it Universal} function of radius, tuned by some galaxy
property $P$,  such as the luminosity or  the   virial  mass that  serves as   galaxy identifier. 
The radius can be measured in three  coordinate systems $R, R/R_D,
R/R_{vir}$:   the physical  coordinate, and  coordinates  focused on the  luminous and the  dark matter distributions
respectively.  
   
\begin{figure}[h]
 \vskip -0.4truecm
\centerline{\psfig{figure=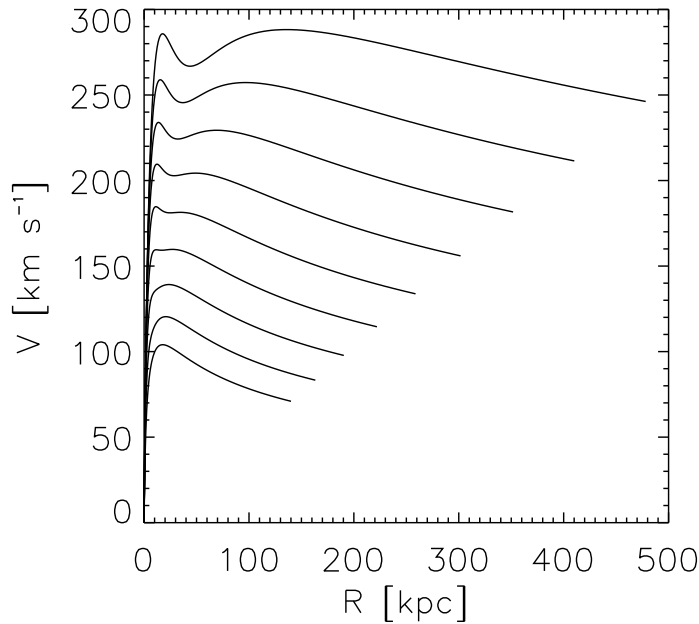,height=9.5cm}}
 \vskip -0.2truecm
 \caption{The Universal Rotation Curve with the  radial coordinate in physical units. 
 Each curve corresponds to $M_{vir}=10^{11}  10^{n/5} \rm M_\odot $, with $n =
1 \ldots 9$ from the lowest to the highest curve. } 
\vskip -0.1cm
 \end{figure}
 
The data used  to build the URC are 1)   616 RCs that  are first arranged  
  in 11 luminosity intervals spanning the whole $I$-band luminosity range $-16.3 < {\rm M}_I< -23.4$,
  with each luminosity  bin  having  $\sim 1500$ velocity measurements, 
  and then  coadded  by arranging  the latter in  radial bins of size $0.3 \,R_D $. 
  This builds  11 synthetic curves $V_{coadd}(R/R_{opt}, M_I)$  out to  $\sim 4 R_D$ (see PSS),
 free  from most of the observational errors and  non-axisymmetric
disturbances present in  individual RCs. These RC's  result very  smooth and  with a  very small intrinsic variance,  but showing a very strong luminosity dependence (see also Catinella et al, 2006) b) the empirical relationship between  RC slope at 2 $R_{opt} $ and  $log \  V_{opt}$ (see PSS)    c) the halo virial velocity $V_{vir}\equiv (G M_{vir}/R_{vir})^{1/2}$, obtained    
from the disk mass vs.  virial  mass relationship  (see Shankar et al. , 2006).

The  {\it URC paradigm},   i.e. the idea according to which halo mass (or the disk mass), eventually   involving   few other quantities, "determines" at   any radii the circular velocity of any spiral within an error that is typically  much  smaller than the true   variations that the RC shows in galaxies and among galaxies is implemented by  the  URC {\it function},   an   analytical    Curve we construct  as the    sum in quadrature of the   disk  and halo contributions to the circular velocity, meant to be the   observational counterpart  of  the NFW velocity profile.

 Then,  we set: $V^2_{URC} = V^2_{URCD} + V^2_{URCH}$. From (1.2) the disk term is:
  
  \begin{equation}
  V^2_{URCD}(x) = {1\over 2} {GM_D\over {R_D}} (3.2 x)^2 (I_0K_0-I_1K_1)             
  \end{equation}

where $x=R/R_{opt}$ and $I_n$ and $K_n$ are the modified Bessel functions computed at $1.6~x $.

\begin{figure}
\vskip -.8truecm
\centerline{\psfig{figure=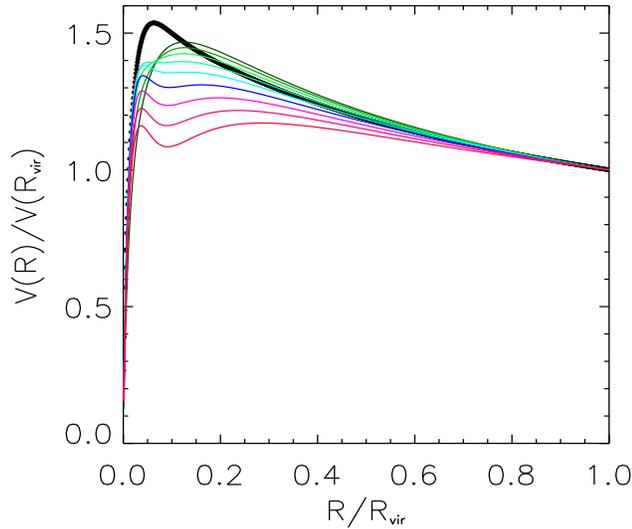,height=8.1cm,width=8.8cm }} 
\vskip-0.4truecm 
\caption{The Universal Rotation Curve, normalized at its
virial value   as a function of normalized "dark"
radius $ R/R_{vir}$. Each curve,  from the highest  to the
lowest, corresponds to  $M_{vir}$   as in Fig. 2. The bold
line is the NFW velocity profile. } 
 \vskip -0.2cm
\end{figure}

For the  DM term we assume  that  the dark halo    follows  a Salucci \& Burkert
(2000) profile,   a cored distribution that can  converges to a NFW one at  outer  radii
 \begin{equation}
\rho (R)={\rho_0\, r_0^3 \over (R+r_0)\,(R^2+r_0^2)}~, 
  \end{equation}
  
$r_0$ is the core radius and $\rho_0$ the    central density
density.  Then:
  \begin{equation}
V^2_{URCH}(R)= 6.4 \ G \ {\rho_0 r_0^3\over R} \Big\{ ln \Big( 1 +
\frac{R}{r_0} \Big) - \tan^{-1} \Big( \frac{R}{r_0} \Big) +{1\over
{2}} ln \Big[ 1 +\Big(\frac{R}{r_0} \Big)^2 \Big] \Big\}~.  
  \end{equation}

Then,  the  URC  function,  has three free parameters   $\rho_0$,  $r_0$,  $M_D$  that are obtained from fitting   $V_{coadd}$ and the other data specified above.

\begin{figure}
 
\centerline{\psfig{figure=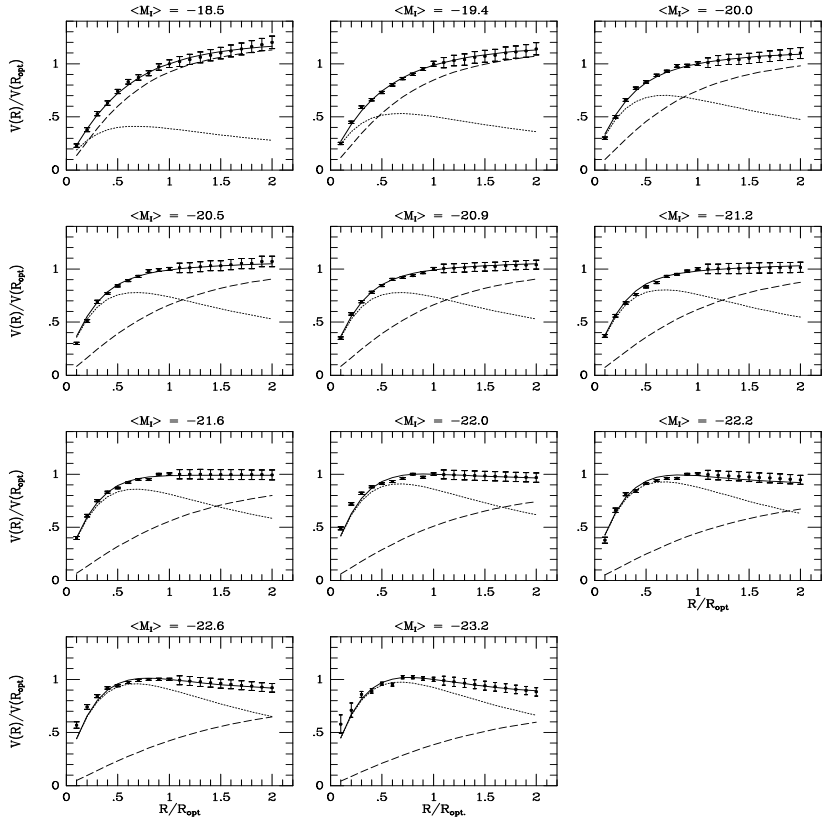,height=12cm,width=12cm}} 
\vskip -0.2cm
\caption {Best  disk-halo  fits to the Universal Rotation Curve (dotted/dashed line: disc/halo)}
\vskip -0.1cm
\end {figure}

In Fig. 3 we show $V_{URC}(R; M_{vir})$, the URC  with the radius expressed in physical units and the objects identified by the halo virial mass; each line
refers to a given halo mass in the range $10^{11}\,  M_{\odot} - 
 10^{13}\,  M_{\odot}$;  the halo mass determines both
the amplitude and the shape of the curve. Note the contribution of
the baryonic component, negligible for small masses but increasingly
important in the larger structures. In Fig. 4  we frame the URC from a full DM perspective by plotting
$V_{URC}(R/R_{vir}; M_{vir})$,  $R/R_{vir}$ is  the radial "dark"
coordinate, and the URC is  normalized   by $V_{vir} \propto
M_{vir}^{1/3}$.

The URC shows  that  the DM halos and stellar  disks  are both  self-similar, but  the whole system is not,
likely  due to the   baryons collapse that has  have broken it in   the innermost $30\%$ of
the halo size.

Let us notice that the RC's of spirals  are {\it critically}  not flat:  their RC slopes     
take all of sort  of values from that of a solid-body  system (i.e. +1)  to   that  of  an  almost Newtonian point-mass (i.e. - 1/2).   
The maximum of the RC occurs at very different radii, for galaxies
of different mass,  viz. at $\simeq 2 R_D$ for the most massive objects and at $\sim 10 R_D$ for the least
massive ones.

 \section{The mass distribution in  Spirals} 
 
The existence of  systematical properties of the  mass distribution in  spirals  was first     claimed by   Persic and Salucci (1988)  and then   successively  confirmed  by independent works (Broeils, 1992; PSS; Rhee, 1997; Swaters, 1999). However, until now  it  has been  considered   by current  theory of galaxy formation only   in a limited way. Likely,  this  is a  consequence of  the theoretical prejudice   that the cosmological galaxy formation process  did not  leave    relevant features  in the   distribution of matter in galaxies  and of  the difficulties   that    simulations have   in  "reproducing" reasonable disk systems.    However,  after Salucci et al  (2007), we believe    that,  in order  to understand the whole process of their  formation   we must take into account  the rich  scenario of the  dark-luminous  interplay occurred in galaxies.     
      
\begin{figure}[h]
 \vskip -0.2cm
\centerline{\psfig{figure=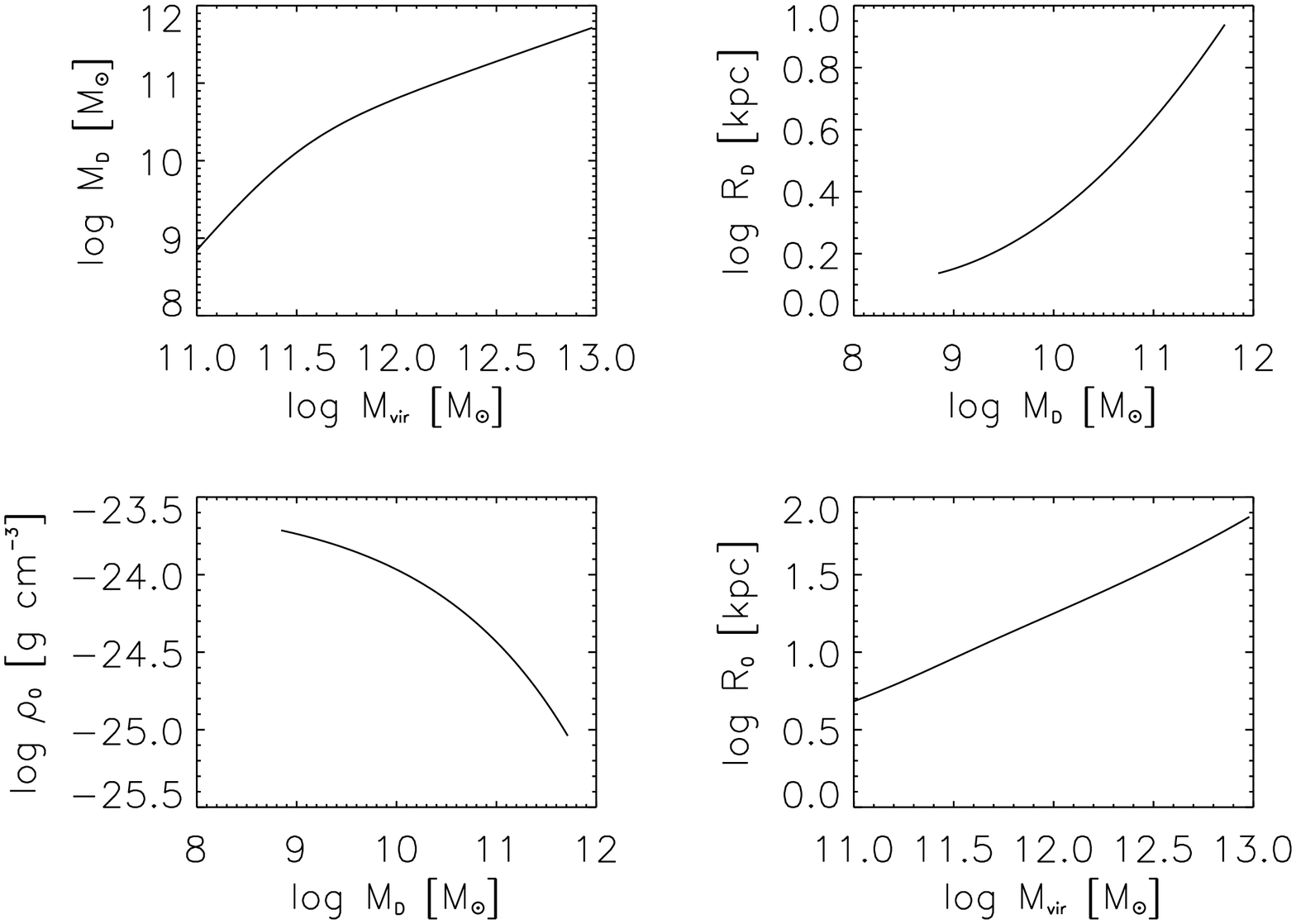,height=9.cm,width=11.5cm }} 
\vskip -0.5truecm 

\caption{The structural parameters of the mass distribution} 
\vskip -0.1truecm
\end{figure}
 
The mass distribution in  Spirals   is obtained  by mass modelling   two very different and complementary  kinematical set of data {\it   a)}   a large number of     {\it individual}   RC's  of  objects  of different luminosity and  {\it b)}  the    URC (see Fig. 5),  that, let us remind,   originates from {\it coadded } RC's;  noticeably,   the  results  obtained from these two different sets of data are  very similar,   indicating so  that they are   robust and reliable. A clear  scenario   of  the mass distribution emerges, see Fig. 6: 
  
$\bullet$ The stellar disk  dominates  the galaxy's inner region  out to the  radius $R_{IBD} $ at which     the DM halo contribution  starts to take over the stellar one. This   sets  the properties of the  Radial Tully Fisher relation  and yields to  the paradigm of the Inner Baryon Dominance:   the inner portions of the  (observed) RC that can   be     accounted  by the stellar matter alone  are  indeed   saturated  by this component.

$\bullet$ At any radii, smaller lower luminosity  objects     have more  progressively more proportion of dark matter i.e. a larger  dark-to-stellar mass ratio.    In detail,  the disk mass  is  $\propto M_{vir}^2$ at small virial  masses, $M_{vir} = 10^{11}M_\odot$    and  $\propto M_{vir}$ at larger masses, $M_{vir}= 10^{13} M_\odot$.   The baryonic  fraction is always  much smaller  than  $\Omega_b/\Omega_{matter} \simeq 1/6  $,  i.e. the cosmological value  and it ranges  between $7\times 10^{-3}$   to   $5\times 10^{-2}$ in line with  is the wellknown evidence that SN  explosions    have removed (or made never condense)  a very large fraction of the original HI material.
  
$\bullet$ Smaller spirals are denser, with the central density spanning 2 order of magnitudes over the mass sequence of spirals.  If this reflects the background density at the formation time,  it   indicates that the  present day population of spirals  has been   hierarchically formed from  $z=5$ to $z=1$,  and,  since then, it has experienced very little merging. 
    
$\bullet$ The  structural parameters of the mass distribution, $\rho_0$, $M_D$, $M_h$, $r_0$  are  remarkably all  related,  see figure (4) and fig (11) of PSS. Notice that, although this evidence  is  connected with the URC paradigm, however      it is not  directly implied by it  nor it implies it. 
  
$\bullet$ The stellar mass-to-light ratio  is  found to lie between 0.5 and 4. The values of  disk masses  derived as above 
agree very well with those obtained by fitting their SED with spectro-photometric models (Salucci, Yegorova, Drory, 2007)  

$\bullet$  The HI component  is  almost always below  the kinematical detectably. However, in low mass  systems it cannot be neglected in the baryonic budget since it is more prominent than the stellar disk.

 \section{The core-cusp issue}
 
A fundamental prediction of the cosmological ($\Lambda$) Cold Dark Matter   simulations is
that virialized dark matter halos  have  an universal spherically averaged density profile 
$\rho_{CDM} (R)$  (Navarro, Frenk \& White, 1997)

\begin{equation}
\rho_{CDM}(R) = \frac{\rho_s}{(R/r_s)(1+R/r_s)^2}
\end{equation}
 where $\rho_s$ and $r_s$ are   strongly correlated  (e.g. Wechsler et al, . 2002),  
  we have:  
$r_{\rm s} \simeq 8.8 \left( \frac{M_{vir}}{10^{11}{\rm M}_{\odot}}   
\right)^{0.46} {\rm kpc}    
$,     
that sets the region of the inner cusp.  The above profile  converges,   at small radii, to a power-law  of index -1 ,  although,     according to   higher resolution simulations,   the actual    value of the index   reaches about    -1.35 (e.g. Moore et al 1999,  Navarro et al. 2004).  

The CDM predictions can be  confronted with observations:  RC's   of disk galaxies, in fact,  probe
the crucial region $0.1 r_s \leq r \leq 2 r_s$.
Flores \& Primack (1994), Moore (1994) claimed,     for some dwarfs,   a  tension between  the kinematical data and the  predictions  of simulations:   DM halos  seemed  to prefer cored density distributions  rather than  cuspy ones (see also  van den Bosch and  Swaters,   2001;  Weldrake et al.  2003).

\begin{figure}[h]
\vskip 0.2cm
\centerline{\psfig{figure=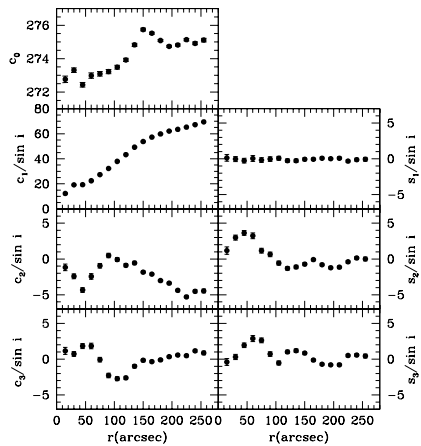,width=5.6cm}}
   \caption{Harmonic decomposition of the 2D RC of DDO 47} 
 \skip 0.2cm  
\end{figure}

The importance of the issue,       that concerns the very  nature of dark matter, 
and the fact  that these   early  results were questioned on several aspects,   has  triggered  
new investigations characterized  by the study of  few proper  test-cases   but with higher quality 
kinematical data and by means of a  properly devised  analysis (Gentile et al., 2004). 
These  improvements were  necessary:  to obtain   reliable DM  profiles    requires  extended,  regular,  homogeneous  RC's   reliable up to their second derivative  and free from    deviations from  the axial symmetry.   Then, up to now,   few tenths of objects  have qualified to undergo such critics-free  investigation (e.g. the list in Donato et al., 2004; Simon et al.; 2005, Gentile et al., 2005; 2007; de Blok, 2007).

In all  these cases  data and simulations   were found in   plain disagreement
on  {\it three}  different aspects: the best-fit  disk + NFW halo mass model a) fits the RC  poorly and it implies  b)  an implausibly  low stellar mass-to-light ratio  and c)     an unphysical  high halo mass.

\begin{figure}[h]
\vskip 0.1cm
\centerline{\psfig{figure=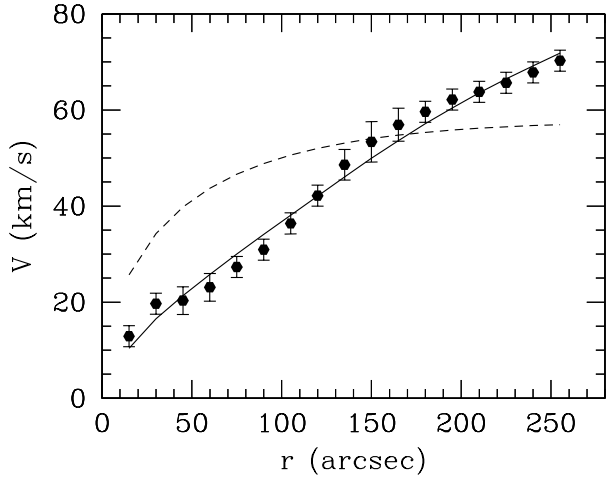,width=8.2cm}}
\caption{DDO 47 rotation curve (filled circles) best-fitted by   Burkert halo + stellar disk (solid line) and by  NFW  halo + stellar disk (dashed line) mass models. The amplitude of non-circular motions is $<5 \ km/s $ }
\vskip 0.1cm
\end{figure}

As an example, it is worth to  discuss  in detail the case of the nearby dwarf  galaxy DDO~47 (Gentile et al 2005).  The  HI observations   have  adequate resolution and sensitivity,    showing  that
the  HI 2-D kinematics   is very regular, with a well-behaved
velocity field.   The observed velocity
along the line of sight $V_{\rm los}$  has been decomposed  in terms of harmonic coefficients:  
$
V_{\rm los} = c_0 + \sum_{j=1}^n [c_j {\rm cos}(j \psi) + s_j {\rm sin}(j \psi)]
$
where $\psi$ is the azimuthal angle,    $c_0 $  is the systemic velocity, 
$c_1$ is the rotation velocity (see Schoenmakers et al.,  1997); it is found that  the   coefficients   $s_1$, $s_3$  $j_2$ 
 have a  small amplitude (see Fig. 7) that excludes    significant  global elongation and lopsidedness of the
potential and detects  non-circular motions    
  with  amplitude and radial profile very   different from that  necessary 
to  hide a cuspy density distribution in the observed rotation curve.  The  RC mass modelling,   shown in Fig. (8),
finds  that the  DDO 47   dark halo  has  a core  radius  of about 7 kpc  and a central density $\rho_0 = 1.4 \times 10^{-24}$
g cm$^{-3}$, i.e. a {\it much} shallower distribution  than that predicted by the  NFW profile.

In all cases studied up to date   a  serious data-prediction  discrepancy  emerges, that becomes  definitive  when we  remind that the actual $\Lambda CDM$  halo profiles  are  steeper than  the standard NFW ones  considered here  and that the   baryonic  adiabatic collapse  has  likely  contracted them further.       
On the other hand, this discrepancy  found in a relatively small number of 
objects  cannot  be extended  to the {\it whole} population of spirals, which is   still not sufficiently investigated.
Let us  stress  that there is not shortage of proposals    to  explain the "density  core  phenomenon"  within  the $\Lambda$CDM scenario itself (e.g Tonini et al 2007).

\section{Conclusions}

The distribution of luminous and dark  matter in galaxies shows amazing properties and a   remarkable systematics  that make it 
as one of the hottest cosmological issues.  There is no doubt that this emerging observational scenario   will be decisive 
in guiding  how  the $\Lambda CDM$-based    theory of galaxy formation must evolve to meet the challenge that the observational data are  posing.     
\label{sec:concl}

\end{document}